\documentstyle[12pt]{article}

\begin{document}

\LaTeX{}\bigskip\ \bigskip\ \bigskip\ 

\begin{center}
Some possible astrophysical applications of diamond anvil cells\bigskip\ 
\bigskip\ 

Vladan Celebonovic\medskip\ 

Institute of Physics,Pregrevica 118,11080 Zemun-Beograd,Yugoslavia\medskip\ 

vladan@phy.bg.ac.yu

vcelebonovic@sezampro.yu

\bigskip\ 
\end{center}

Abstract: This is a review of basic methods of studies of materials under
high static pressure in diamond anvil cells,and some applications of these
methods to astrophysically relevant materials.As the paper was originally
published in 1993.,an addendum gives a hint of some recent developements.
\newpage\ 

The pioneering experiments of Bridgman,in the early and middle parts of our
century,have marked the beginning of the modern period of experimental
high-pressure work ( for a review of his work see Bridgman,1964).Bridgman
used large volume presses which could contain large samples,and in which P-T
gradients were diminished,but which had the disadvantage of a limited P-T
range,no direct observation of the sample was possible,and they were
expensive to install and maintain in operating conditions.

A\ breaktrough in high pressure experimental techniques occured near the
middle of this century (Lawson and Tang,1950;Jamieson et al,1959;Weir et
al.,1959) with the invention of the diamond anvil cell (DAC).The first DACs
were built with the aim of performing high pressure x-ray diffraction
studies and infrared absorption measurements under high pressure.Later
evolution,described in detail in the literature ( such as
Jayaraman,1983,1986;Willi\qquad ams and Jeanloz,1991;Angel et al.,1992;Itie
1992) has converted the DAC into a versatile quantitative tool for physical
research.

The basic principle of the DAC is extremely simple.A sample is placed
between the flat paralel surfaces of two oposed diamond anvils,and it is
subjected to pressure when the diamonds are pushd together by an external
force.Variations in DAC types arise from different ways of generating the
external force,transmitting it to the daimonds and aligning them.In order to
achieve hydrostatic experimental conditions,a gasket is inserted between the
diamonds.

The gasket is a thin metal foil,with a hole containing the pressure
transmitting medium and the sample in its center.Pressure is measured in the
''ruby scale'': the R lines of ruby ($Al_2O_3:Cr^{3+})$ have a well known
pressure shift.Accordingly,a small chip of ruby is placed in the hole in the
gasket,and its fluorescence is exxcited by a laser or any other source of
strong light.This scale is linear up to at least 30 GPa (Jayaraman,1983) ;
in its non-linear form the ruby scale can be applied for pressure
measurements up to 250 GPa \qquad \negthinspace ( Ruoff, 1992a).At higher
values of pressure,only X-ray diffraction measurements can be
performed.Experiments in DACs are complicated by the miniaturized scale;for
example,the hole containing the sample and a chip of ruby has a diameter of
only 200$\mu m$ ,while the typical size of the samples is of the order of
40-50 $\mu m$.Experiments can be performed in the interval of temperatures
between 4K and around 7000 K ( Williams and Jeanloz,1991).\newpage\ 

What applications can DAC experiments have in astrophysics?It is a ''fact of
life'' that no direct observation of planetary or satellite interior is
possible.Some of the observable planetary parameters critically depend on
the conditions prevealing in their interiors,and the only experimental
method for investigating them is the use of DACs.

For example,the giant planets of the Solar System contain a large percentage
of hydrogen.The obvious question is how does it behave under extremely high
pressure ( of the order of hundreds of gigapascals,as in the center of
Jupiter ).Theory predicts that hydrogen becomes metallic at a pressure of
200-300 GPa (for example, Wigner and Huntington,1935;Barbee et al.,1989;
Ashcroft,1989 and many other papers).Claims were recently made that
metallization of hydrogen has been detected in a DAC at a pressure of $%
P\cong 200GPa$ (Mao and Hemley,1989),but they were later shown to be
incorrect (Ruoff,Greene,Ghandehari and Xia,1992 ). Accordingly,the existence
of metallic hydrogen in deep interiors of the giant planets,and its possible
consequences on the observable planetary parameters is still an unsettled
question.

A closely related problem is the behaviour of ice under high pressure.A new
high pressure,low temperature phase,called ice XII has recently been
discovered ( Bizhigitov and Sirota,1986).It is stable for temperatures in
the range between 90 and 250 K,and in the pressure interval 1200 - 2150
MPa.The preceeding phase, ice XI,becomes metallic at P = 1.76 TPa ( Hama et
al.,1990).Such data are relevant for modellization ( and interpretation of
obseervations ) of the giant planets and their satellites; to the author's
knowledge,they have not been widely used.

What about the interior of the Earth/Its composition is one of themost
important planetological problems (Knittle and Jeanloz,1991b).It is
generally assumed that it consists of a metallic core surrounded by a rocky
crust \qquad \qquad ( Jeanloz,1990 ) .\newpage\ Experiments in DACs have
given valuable indications about the conditions in its deep interior,such as
the central temperature ( Williams,Jeanloz,
Bass et al.,1987 ),the temperature at the core-mantle boundary ( Knittle 
and Jeanloz,1991b ),or the possibility of chemical reactions between the 
silicates and liquid iron ( Knittle and Jeanloz,1991a ).An analysis of 
melting of iron under high pressure ( Celebonovic,1993a ) has given 
indications about the changes of the Gruneisen parameter of iron under high 
pressure,which is an example of a planetologically motivated result in solid-
state physics.Numerous other examples of DAC experiments giving planetologically 
interesting information can be found in the literature ( such as Jeanloz ,1989,
1990; Williams and Jeanloz,1991 ).

Instead of a conclusion,an information: high pressure studies,both
experimental and theoretical, are going on in the Institute of Physics.Those
interested in any possible kind of a contact and/or collaboration are
wellcome to contact the author at either of the email addresses indicated.%
\bigskip\ 

Addendum ( December 20,1998 )\medskip\ 

This mini review was written in 1993.High pressure studies have advanced
since then,and it would be over ambitious to attempt reviewing the whole of
it in one paper.However,two notes are in order:

Metallization of hydrogen is still a problem.Recent work by the group of
Prof.Ruoff at Cornell University indicated that metallization of hydrogen
under static pressure does not occur for $P\leq 342$ GPa (Nature,issue dated
May 7,1998).This result poses serious problems for theoretical
studies,because it oposes predictions of the metallization pressure of
hydrogen made during the last 60 years. On the other hand,work at the
Lawrence Livermore National Laboratory on fluid hydrogen has shown a
transition from a semiconducting to a metallic liquid at $P\cong
140GPa,T\cong 3000K$ (S.T.Weir,A.C.Mitchell and W.J.Nellis,Phys.Rev.Lett.,%
{\bf 76},1860 (1996)).

Although the ruby scale is widely applied,work is going on with the aim of
finding new pressure sensors.A recent example is the work by Jovanic et al (
B.R.Jovanic,B.Radenkovic,Lj.D.Zekovic,J.Phys.:Condens.Matter,{\bf 8},4107
(1996)) where they have investigated the transition $^5D_0\rightarrow ^7F_2$
in $Y_{1.9}Eu_{0.1}O_3$ and obtained the pressure dependence of the
fluorescence lifetime for this transition.

\newpage\ 

\begin{center}
References\medskip\ 
\end{center}

Angel,R.J.,Ross,N.L.,Wood,I.G. and Woods,P.A.:

1992,Phase Transitions,{\bf 39}, 13.

Ashcroft,N.W.: 1989,Nature,{\bf 340}, 345.

Barbee,T.W.,Garcia,A and Cohen,M.L.:

1989,Nature,{\bf 340},369.

Bizhigitov,T.B.and Sirota,N.N.: 1986,JETP Letters,{\bf 44},417.

Bridgman,P.W.: 1964,Collected Experimental Papers ,Vol. I-VII,

Harvard University Press,Cambridge,USA.

Celebonovic,V.: 1993a,Earth,Moon and Planets,{\bf 61},39 .

Hama,J.,Shiomi,Y and Suito,K.:

1990,J.Phys.:Cond.Matt.,{\bf 2},8107.

Itie,J.P.: 1992,Phase Transitions,{\bf 39},81.

Jamieson,J.C.,Lawson,A.W.and Nachtrieb,N.D.:

1959.Rev.Sci.Instr.,{\bf 30},1016.

Jayaraman,A.:1983,Rev.Mod.Phys., {\bf 55},65.

Jayaraman,A.:1986,Rev.Sci.Instr.,{\bf 57},1013.

Jeanloz,R..:1989,in:Proc.of the Gibbs Symposium,Yale Univ.,

May 15.-17.,1989.,p.211,ed by the Amer.Math.Soc.,Boston.

Jeanloz,R.:1990,Ann.Rev.Earth Planet.Sci.,{\bf 18},357.

Knittle,E.and Jeanloz,R.:1991a,Science,{\bf 251},1438.

Knittle,E.and Jeanloz,R.: 1991b,J.Geophys.Res.,{\bf 96B},16169.

Lawson,A.W.and Tang,T.Y.:1950,Rev.Sci.Instr.,{\bf 21},815.

Mao,H.K.and Hemley,R.J.:1989,Science,{\bf 244},1462.

Ruoff,A.L:1992a,in: Ann.Techn.Rep.,Mater.Sci.Ctr.,Cornell

Univ.,July 1,1991.-June 30,1992.

Ruoff,A.L.,Greene,R.G.,Ghandehari,K.,and Xia,Q.: 1992,

Cornell Univ.Mat,Sci.Center.Rep.,MSC \#7439.\newpage\ 

Weir,C.E.Lippincot,E.R.,Van Valkenburg,A.and Bunting,E.N.:

1959,J.Res.Natl.Bur.Stand.Sec.{\bf A63},55.

Williams,Q.,Jeanloz,R.,Bass,J.Svendsen,B.,

Ahrens,T.J.:1987,Science,{\bf 236},181.

Williams,Q.and Jeanloz,R.:1991,in:Molten Salt Techniques,vol.4,p.193,

ed by.R.J.Gale and D.G.Lovering,Plenum Publ.Corp.,London.

Wigner,E.and Huntington,H.B.: 1935,J.Chem.Phys.,{\bf 3},764.

\end{document}